\documentclass[%
 aps,
 jmp,%
 amsmath,Amssymb,
 reprint
]{revtex4-1}

\usepackage{graphicx}
\usepackage{dcolumn}
\usepackage{bm}
\usepackage{mathrsfs}
\usepackage{cancel}
\usepackage{float}
\usepackage{textcomp}
\usepackage[normalem]{ulem}
\usepackage{verbatim} 

\begin{document}

\title[ ]{Energy and RF Cavity Phase Symmetry Enforcement in Multiturn Energy Recovery Linac Models}
\author{R. Koscica}
\author{N. Banerjee}
\author{G. H. Hoffstaetter}
\author{W. Lou}
\author{G. Premawardhana}
\affiliation{CLASSE, Cornell University, Ithaca, NY 14853, USA}

\date{April 2019}

\begin{abstract}
 In a multi-pass Energy Recovery Linac (ERL), each cavity must regain all energy expended from beam acceleration during beam deceleration, and the beam should achieve specific energy targets during each loop that returns it to the linac. For full energy recovery, and for every returning beam to meet loop energy requirements, we must specify and maintain the phase and voltage of cavity fields in addition to selecting adequate flight times. These parameters are found with a full scale numerical optimization program. If we impose symmetry in time and energy during acceleration and deceleration, fewer parameters are needed, simplifying the optimization. As an example, we present symmetric models of the Cornell BNL ERL Test Accelerator (CBETA) with solutions that satisfy the optimization targets of loop energy and zero cavity loading. An identical cavity design and nearly uniform linac layout make CBETA a potential candidate for symmetric operation. 
\end{abstract}

\maketitle

\section{Introduction} 
The Energy Recovery Linac (ERL) was first proposed by Tigner in 1965 as an economically efficient accelerator capable of producing particle beams of high quality, high current, and low cross-sectional area \cite{Tigner1965}. An ERL reclaims energy from decelerating particle bunches to reduce the net power consumption of the accelerating radio frequency (RF) cavities. Return loops connect the exit end of the linac to the entrance, allowing the beam to recirculate through multiple accelerating passes of the linac. After the highest energy target is achieved, the beam decelerates through another series of linac encounters. In an ERL designed to use common recovery transport, the beam traverses the same physical linac and return loops during acceleration and deceleration. The energy of a decelerating beam returns to the RF cavities, which reuse this energy to accelerate future particle beams \cite{erl}. The beam must satisfy two conditions for full energy recovery:
\begin{itemize}
 \item \textit{Energy recovery.} During deceleration, each cavity should regain the same amount of energy that it transferred to the beam during acceleration. If satisfied, the power load on each cavity from the beam will on average be zero in steady state.
 \item \textit{Reasonable energy targets.} If the beam is intended for experimental applications, it must achieve the desired energy target. Additionally, the energy of the beam during each return loop must satisfy design criteria for the particular ERL construction.
\end{itemize} 
Due to the varying longitudinal velocity ($v$) of the beam throughout the ERL, it is challenging to find RF cavity phases that provide the desired acceleration during all linac passes. Adjustment of loop length alone may not guarantee the appropriate energy recovery in all cavities. If all cavity phases are adjusted to maximal energy gain for the highest energy particles, where $v \approx c$, then low-energy beams will experience RF phases slipped away from maximal energy gain, and synchrotron radiation in the return loops can cause additional offsets in beam time of flights \cite{erl}. The individual phases and voltage settings of each RF cavity, as well as the time of flight through the return loops, are the parameters for optimizing cavity loads and the beam energies.\par
\begin{figure}
\includegraphics[width=\linewidth]{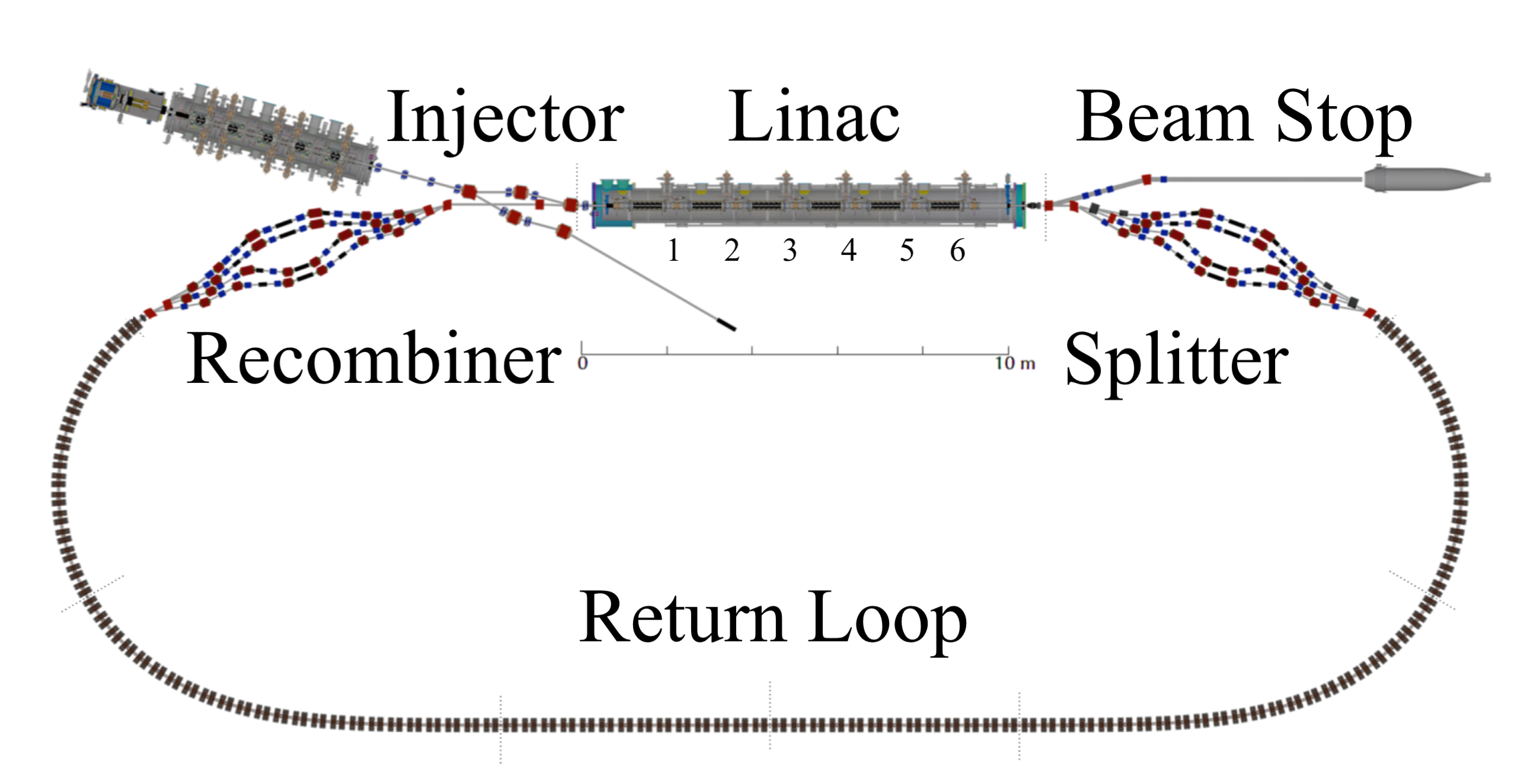}
\caption{Layout of the CBETA ERL \cite{designreport}. A 6~MeV beam is generated in the injector, accelerates through a linac with six 7-cell cavities, circulates clockwise through the return loop, and recirculates to cover a total of 4 accelerating and 4 decelerating linac passes. The particle path concludes at the beam stop (top right). Splitter and recombiner regions at either end of the linac independently control the flight times of beams with the target energies of the 4 loops.}
\label{fig:cbeta}
\end{figure}

In this paper, we use a mixture of theory and numerical optimization algorithms and present a time-symmetric method of identifying loop lengths, RF voltage, and RF phase settings. This symmetry reduces the number of fit parameters required during optimization. During ERL operation, it is possible for the symmetry to be intentionally broken due to energy extraction from use of the highest energy beam; such an effect is not considered in this study, which solely examines an ERL in steady state. We use a model of the Cornell BNL ERL Test Accelerator (CBETA), which is a common transport ERL with 4 physically distinct return loops and a linac that holds 6 evenly spaced accelerating cavities (Fig.~\ref{fig:cbeta}). 

CBETA brings a 6~MeV injected beam to 150~MeV over 4 accelerating passes and returns it to 6~MeV over 4 subsequent decelerating passes \cite{designreport}. The 150~MeV beam, which travels around the 4$^{th}$ ERL loop, can be used as a compact synchrotron radiation source or for internal target experiments in nuclear and elementary particle physics \cite{ERLuses}. Afterward, the high-energy beam is recirculated through the full multi-turn path for deceleration. \par 
Suppose that CBETA RF phases and loop lengths are set to give full energy recovery and a maximum energy of 150~MeV for an ultra-relativistic beam ($v=c$), but we inject a transversely on-axis 6~MeV electron beam with a 40~mA current. In a simple thin lens cavity model (to be discussed in more detail in the following sections), the six cavities have positive loads of 46, 40, 38, 37, 38, and 39~kW. However, only 2-4~kW are available for beam acceleration in each cavity in the CBETA main linac \cite{designreport}. In this ultra-relativistic phasing scheme, the beam completes the full ERL circulation with an energy of nearly 12~MeV. Over the 8 passes, the beam also encounters phase slip from the ultra-relativistic case, where the maximum phase slip values at each cavity are 0.55, 0.37, 0.52, 0.37, 0.53, and 0.37~radians. \par 
If the CBETA phases and loop lengths are set for a $v=c$ beam as previously described, but the 40~mA beam is injected with a 12~MeV energy instead of 6~MeV, the six cavities have positive loads of 28, 26, 26, 25, 25, and 26~kW. The maximum phase slip values for each cavity are 0.42, 0.28, 0.40, 0.28, 0.41, and 0.29~radians. From these example loads and phase slips, as well as those of the 6~MeV case, we observe that energy recovery for a non-ultrarelativistic beam is not well achieved by using $v=c$ phases, but the load and phase slip do decrease if a more relativistic beam is used. \par 
The beam power load can be reduced by optimizing the phase and loop length settings for a 6~MeV injected beam. In this study, we develop a symmetric acceleration system with load and beam energy objectives. The system is implemented in CBETA simulations to illustrate the use of symmetry in enhancing the optimization process, and the resulting objectives are calculated in ERL models of increasing complexity. For CBETA, the maximum beam energy is 150~MeV, and synchrotron radiation is not relevant at this energy; we therefore do not consider energy losses due to radiation. If higher energy ERLs are modeled, synchrotron radiation would become relevant. \par
We also generalize our approach to ERLs where the beam passes $M$ times through a single linac of $N$ cavities. Each injected particle bunch results in $M-1$ beams looping back to the linac at different points in time. In a common transport ERL, the beam that has been accelerated by $m \leq \frac{M}{2}$ linac passes has approximately the same energy as the beam that is yet to be decelerated $m$ times; these two beams then traverse the same beam pipe, and one speaks of a $\frac{M}{2}$-turn ERL. \par 
If large energy aperture optics are used, all $M-1$ returned beams can travel in one vacuum pipe. Such an optics configuration can be realized with a Fixed Field Alternating-gradient (FFA) design, as in CBETA  \cite{designreport}. Assuming the beam energy stays constant during the return loop, then the energy changes occur a total of $M\cdot N$ times over all cavity encounters. \par
If the profile of energy gain and loss is symmetric through the entire ERL, then we call this a symmetric ERL. With this symmetry, the energy that the beam gains when it traverses a cavity for the $i^{th}$ time is the same as the energy that the beam loses during the $(MN-i+1)^{th}$ time. \par 
Using these notations, CBETA is a 4-turn ERL with $N=6$ and $M=8$. Because of its FFA optics, CBETA only has one vacuum pipe for all 7 returned beams. Additionally, by adjusting parameters appropriately, it can be operated as a symmetric ERL. \par 

\section{Optimization System Parameters} 
Suppose an $\frac{M}{2}$-turn, single-linac ERL has $N$ cavities and $M$ linac passes. To indicate individual loops and cavities, we use indices $m$ and $n$ such that $1\leq m \leq M$ and $1 \leq n \leq N$. \par 

\subsection{Objectives}
During ERL operation, beam timing must ensure that an appropriate amount of energy is transferred during each and every cavity passage. Optimal energy recovery requires there to be a minimal load on all RF cavities: 
\begin{equation}\label{eq:eload}
 E_{\text{load},n} = \sum_{m=1}^{M} \Delta E_{mn} \rightarrow 0 \text{~eV}
\end{equation}
where $\Delta E_{mn}$ is the energy gain in cavity $n$ during pass $m$. When optimizing an ERL, we use $E_{\text{load},n}$ as $N$ objective functions that are to be minimized. For the beam to follow the design orbit, the beam energy $E_{\text{loop},m}$ at the end of the $m^{th}$ linac pass must be close to the design energy, $E_{\text{des},m}$, to be compatible with the magnet settings,
\begin{equation}\label{eq:eloop}
 E_{\text{loop},m} = E_{\text{loop},0} + \sum_{k=1}^{m} \sum_{n=1}^{N} \Delta E_{kn} \rightarrow E_{\text{des},m}
\end{equation}
where $E_{\text{loop},0}$ is the beam energy at initial injection. An optimization then minimizes the objective functions $|E_{\text{loop},m}-E_{\text{des},m}|$ for $1\leq m\leq ( M-1)$. \par
In a perfect ERL where $E_{\text{load},n} = 0$ for all $n$, Eq.~(\ref{eq:eload}) leads to $E_{\text{loop},M} = E_{\text{loop},0}$. Therefore, the objective function for $m=M$ is not used, and we have $N+M-1$ objectives. \par
A symmetric ERL has $E_{\text{des},m} = E_{\text{des},M-m}$. If the operation of the cavities guarantees a symmetric energy profile, then also $E_{\text{loop},m} = E_{\text{loop},M-m}$ during each optimization step, and the beam loading is symmetric as well: $E_{\text{load},n}=E_{\text{load},N-n+1}$. If $N$ is odd, then the central cavity in the linac already has zero load when ERL symmetry exists, and the central load ceases to be a useful objective. Using the Gauss bracket to designate the floor of a real number, we then obtain only [$\frac{N}{2}$]$+\frac{M}{2}$ objectives. If an ERL has return loops that can be designed or adjusted to match an expected beam energy, one can further reduce the number of objectives by not requiring specific target energies in the intermediate loops, and only specifying the highest energy of loop $\frac{M}{2}$. In this case, the loop magnet settings would need to be adjusted according to the post-optimization energies of the intermediate loops. The system then reduces to [$\frac{N}{2}$]+1 objective functions.\par

For CBETA, we therefore generally have 13 objectives. As a symmetric ERL, we have 7 objectives. If intermediate energies are not used as targets, we only have to meet 4 objectives. In the latter scenario, appropriate choices of optimization input parameters will increase the likelihood that the resulting intermediate loop energies fall within the design capabilities of the CBETA splitter and recombiner sections. \par

\subsection{Degrees of Freedom}
To identify degrees of freedom that will optimize load and loop energy objectives, we need independent parameters that affect $\Delta E_{mn}$. In general, we can control the following parameters that affect cavity energy gain:
\begin{itemize}
 \item Initial RF phase of the $n^{th}$ cavity, $\phi_{0,n}$. 
 \item Voltage of the $n^{th}$ cavity, $V_n$. 
 \item Travel time through the $m^{th}$ loop, $t_{\text{loop},m}$. 
\end{itemize}
In total, these provide $2N+M-1$ degrees of freedom. \par 
In a symmetric ERL, the last [$\frac{N}{2}$] cavities are operated symmetrically to the first [$\frac{N}{2}$], and therefore do not provide phase or voltage degrees of freedom. If $N$ is odd, then the phase of the central cavity must also be chosen such that it operates symmetrically to itself; however, the voltage of this cavity is still free. Together, the [$\frac{N+1}{2}$] voltages and [$\frac{N}{2}$] phases yield $N$ degrees of freedom from the cavities. The energy symmetry also makes the $(M-m)^{th}$ time of flight equal to the $m^{th}$ one. There are then $N+\frac{M}{2}$ degrees of freedom. \par 
In the case of CBETA, we generally find 19 degrees of freedom. With fully symmetric operation, we have 10 degrees of freedom. \par

\subsection{Efficient Optimization}
In general, we can make the optimization more efficient by having a small number of objectives and equally few degrees of freedom. In each ERL system, there are more degrees of freedom than objectives. To speed up the optimization process, one can decrease the number of degrees of freedom. For example, one can set all cavity voltages to the same value. In a general ERL, there are $N$ more degrees of freedom than constraints; we can therefore make the $N$ voltages constant to achieve an optimization system with an equal number of objectives and degrees of freedom. \par
A symmetric ERL has [$\frac{N+1}{2}$] voltage degrees of freedom, and these can also be set to a constant value. If the voltage of one cavity must be reduced for operational reasons, then that of the symmetric cavity must be reduced as well, and the voltages of the other cavities should be increased. In a reduced ERL with only one loop energy ($E_{\text{loop},\frac{M}{2}}$) to be optimized, we can additionally set the return times of the intermediate loops to fixed values. This removes $\frac{M}{2}-1$ degrees of freedom. The remaining phases and the peak-energy return time lead to an equal number of degrees of freedom and objectives, [$\frac{N}{2}$]+1, as found in Table~\ref{table:freedom-objectives}.\par 
In CBETA, an optimization that only considers the highest loop energy would then use 4 degrees of freedom: 3 cavity phases and 1 time of flight for the $\frac{M}{2}^{th}$ loop. This minimal reduced system for CBETA is found in the lower half of Table~\ref{table:freedom-objectives}. \par

\renewcommand{\arraystretch}{1.3}
\setlength\doublerulesep{2mm}
\begin{table}
\begin{center}
\begin{tabular}{ |c|c|c|c| }
 \hline
General Single- & Full System & Symmetric & Reduced \\
Linac ERL & & & \\
\hline
Objectives        & $N+M-$1 &[$\frac{N}{2}$]+$\frac{M}{2}$&[$\frac{N}{2}$]+1 \\
DoF: all   & $2N+M-$1&N+$\frac{M}{2}$&N+$\frac{M}{2}$\\
DoF: fixed $V_n$       & $N+M-$1 & [$\frac{N}{2}$]+$\frac{M}{2}$&[$\frac{N}{2}$]+$\frac{M}{2}$\\
DoF: fixed $V_n$, $t_\text{loop}$ &$N+$1&[$\frac{N}{2}$]+1&[$\frac{N}{2}$]+1\\
\hline\hline
CBETA & - & - & - \\
\hline
Objectives        &13 & 7 & 4 \\
DoF: all   & 19 & 10 & 10\\
DoF: fixed $V_n$       & 13 & 7 & 7 \\
DoF: fixed $V_n$, $t_{\text{loop}}$& \cancel{ 10 } & \cancel{ 4  } & 4\\
\hline

\end{tabular}
\end{center}
\caption{Objectives and degrees of freedom (DoF) for: a fully independent ERL, one with symmetric operation, and a reduced symmetric system where only the highest loop energy, $E_{\text{loop},\frac{M}{2}}$, is used as an objective function. Also shown are decreased numbers of degrees of freedom when only voltage ($V_n$), or both voltage and intermediate loop flight times ($t_{\text{loop},m\neq \frac{M}{2}}$), are set at constant values. Crossed out options are unhelpful for optimization.}
\label{table:freedom-objectives}
\end{table}
\renewcommand{\arraystretch}{1.}
During optimization, a solution generally cannot be found if there are fewer degrees of freedom than objective functions. For example, the degrees of freedom for the symmetric CBETA system should not be minimized with both constant $V_n$ and constant $t_{\text{loop},m\neq \frac{M}{2}}$, because this will result in using only 4 degrees of freedom to fulfill 7 objectives (Table~\ref{table:freedom-objectives}, bottom row).

\par

\section{ERL Symmetry Conditions}
A symmetric ERL has a significantly reduced number of objective functions and degrees of freedom compared to a non-symmetric ERL, and it can therefore be optimized much more easily. This is true for accelerator simulations as well as for experimentally finding the desired accelerator settings. In this section, we discuss operation conditions for a single-linac, common transport system that create a symmetric ERL. This is achieved most easily if the cavities are arranged with mirror symmetry about the center of the linac cryomodule. This means that the distance from the $n^{th}$ cavity to the next is the same as the distance between the $N-n^{th}$ cavity and its next neighbor. There will also be cavity phases where the standing waves in the $n^{th}$ cavity have a field that is the mirror image of the $(N-n+1)^{th}$ cavity field. Mirror symmetry in the fields always exists if the geometry of the $n^{th}$ cavity is the mirror image of that of the $(N-n+1)^{th}$ one. \par
Most ERLs have linacs constructed from regularly spaced cavities; if these ERLs are also single-linac and use common recovery transport, then they are already nearly symmetric physical systems, even if symmetry is not an explicit design choice. In CBETA, the cavities are designed identically, but they are installed such that all are facing the same direction: an input coupler is located on the downstream end of each cavity. However, because the cavity cells are constructed in a symmetric way, and the field from the input coupler is small compared to the main cavity field, these cavities are symmetric to a sufficient degree. \par
Mirror symmetry in the cavities, as well as over the distance between cavities, can result in a symmetric ERL if we choose the particle timing and cavity phases such that following a particle forward from injection, or tracking it backward in time from the beam stop, both result in exactly the same electric fields seen by the particle. If these conditions are met, the acceleration and deceleration profiles of the forward and backward-traveling particle are then identical, and we have a symmetric ERL, \textit{i.e.} the beam decelerates with the same energy steps as during acceleration. \par 
To produce a symmetric ERL, we need a way to set cavity phases such that accelerating and decelerating cavities can be symmetric. In the following discussion, we first show that symmetry can be established in a linear sequence of two cavities. We then transfer this symmetry to the closest equivalent single-linac ERL: a 2-pass, 1-cavity system, where the same cavity is traversed twice. The ERL is then expanded to an arbitrary even number of $M$ passes and 1 cavity. Finally, we combine the findings of the linear and 1-cavity systems to arrive at the symmetry requirements for a general $M$-pass, $N$-cavity single-linac ERL.

\subsection{Linear Sequence: 2 Cavities}
Consider a straight linac with two independent cavities ($A$, $B$) of frequency $\omega$, arranged in mirror symmetry about their center. 
For $A$ to add as much energy as $B$ removes, seek the input phase $\phi_{\text{in},B}$. 
The field of a standing wave in cavity $A$ depends on longitudinal position $s$, time $t$, and input phase. Between entrance $s=0$ and exit $s=L$ of $A$, the field observed by the particle is, 
\begin{equation}\label{eq:efield-A}
    \mathscr{E}_{A}(s,t)=\mathscr{E}_{A0}(s) \sin\big(\omega (t-t_{\text{in,A}})+\phi_{\text{in},A}\big),
\end{equation}
where the particle enters $A$'s field at the input time $t=t_{\text{in},A}$. The input phase $\phi_{\text{in},A}$ is independent of particle entrance energy, but it can be expressed in an energy-dependent form for particle speed $v\leq c$,
\begin{equation}\label{eq:phiin-gen}
    \phi_{\text{in},A} = \phi_{\text{in},A}^v + \hat\phi_A^v = \phi_{\text{in},A}^c + \hat\phi^c,
\end{equation}
where the relative input phase, $\phi_{\text{in},A}^v=0$, is the parameter controlled during ERL operation. The phase offset $\hat\phi_A^v$ enables the particle with speed $v$ to experience maximum acceleration at an on-crest phase of $\phi_{\text{in},A}^v=0$. In general, $\hat\phi_A^v \neq \hat\phi_B^v$ because $v_A \neq v_B$. \par 
By convention, the spatial RF field dependence $\mathscr{E}_{A0}(s)$ is chosen to start with a positive value in the first cell. Due to the symmetry or anti-symmetry of this spatial function, $\mathscr{E}_{B0}(L-s) = \pm\mathscr{E}_{A0}(s)$, where the sign ($+$) is for odd and ($-$) is for even numbers of cells per cavity. \par 


The exit of $B$ is a distance $L_s$ from the entrance of $A$, and $\mathscr{E}$ describes the net field that the design particle experiences from both $A$ and $B$. The particle reaches $s$ at time $t(s)$. Opposite fields in $A$ and $B$ therefore occur if,

\begin{subequations}
\begin{align}
    \mathscr{E}(L_s-s) &= -\mathscr{E}(s)\label{eq:efield-equalbutopposite}\\
    \mathscr{E}_B(L-s,t(L_s-s)) &= -\mathscr{E}_A(s,t(s)).\label{eq:efield-equalbutopposite-ab}
\end{align}
\end{subequations}
At these positions, the time spent in $B$ or remaining in $A$ must also be equivalent,
\begin{equation}\label{eq:efield-timesym}
t(L_s-s)-t_{\text{in,B}} = T_A - (t(s)- t_{\text{in,A}}),
\end{equation}
where $T_A$ is the full duration that the particle spends in $A$. We then equate the mirror symmetric field, $\mathscr{E}_B(L-s,t(L_s-s))$, with the cavity $A$ field in Eq.~(\ref{eq:efield-A}) using the spatial relation from Eq.~(\ref{eq:efield-equalbutopposite-ab}) and time from Eq.~(\ref{eq:efield-timesym}),
\begin{equation}\label{eq:efield-B}
\begin{aligned}
    \mathscr{E}_{B}&(L-s,t(L_s-s))\\ =&\pm\mathscr{E}_{A0}(s) \sin(\omega (T_A - t(s) + t_{\text{in,A}}) +\phi_{\text{in},B})\\
    =& -\mathscr{E}_{A0}(s) \sin(\omega (t(s)-t_{\text{in,A}})+\phi_{\text{in},A}).
\end{aligned}
\end{equation}
Eq.~(\ref{eq:efield-B}) leads to $\phi_{\text{in},B}$ for either odd or even-cell cavities, using $\phi_{\text{out},A} = \phi_{\text{in},A}+\omega T_A$,
\begin{equation} \label{eq:inputphase}
\begin{aligned}
   \phi_{\text{in},B} &=  - \phi_{\text{out},A} = - \phi_{\text{in},A}-\omega T_A & [\text{odd}]\\
   \phi_{\text{in},B} &=  \pi - \phi_{\text{out},A} = \pi - \phi_{\text{in},A} - \omega T_A. & [\text{even}]
\end{aligned}
\end{equation}
With this relation, traveling backward in cavity $B$ yields the same time-dependent fields as traveling forward in $A$, and $B$ removes the same amount of energy that $A$ adds.  \par

Suppose that $A$ and $B$ are pillbox resonators operating in the fundamental mode with $\beta=1$, where one pillbox is traversed in half an oscillation by a particle with $v=c$. The energy gain of an ultra-relativistic particle is,
\begin{equation}\label{eq:deltaE-A}
 \Delta E_A =  qV_A\cos(\phi_{\text{in},A}^c),
\end{equation}
where $q$ is the charge and $V_A=V_B$ is cavity voltage \cite{rfbasics}. \par 
If $A$ and $B$ are each series of adjacent pillboxes in a multi-cell $\pi$-mode cavity, then the ultra-relativistic particle crosses cavity $A$ with a time of flight $\omega T_A = n_\text{cell} \pi $, where $n_\text{cell}$ is the odd or even number of cells per cavity. To find the energy change from $B$, apply the Eq.~(\ref{eq:inputphase}) symmetry conditions to the input phase of $B$,
\begin{align*}
 \Delta E_{B} &=  qV_B\cos(-\phi_{\text{out},A}^c) &[\text{odd}]\\
        &= qV_A \cos(\omega T_A + \phi_{\text{in},A}^c)\\
        &= - qV_A \cos(\phi_{\text{in},A}^c) = -  \Delta E_{A}\\
 \Delta E_{B} &=  qV_B\cos(\pi-\phi_{\text{out},A}^c) &[\text{even}]\\
        &= qV_A \cos(\pi - \omega T_A - \phi_{\text{in},A}^c)\\
        &= - qV_A \cos(\phi_{\text{in},A}^c) = -  \Delta E_{A}.
 \end{align*}
This verifies the Eq.~(\ref{eq:inputphase}) symmetry conditions: $B$ removes as much energy as $A$ adds.\par
We now search for a way to establish similar symmetry in an ERL. \par 

\subsection{ERL: 1 Cavity, 2 Passes}
Consider an ERL with the minimum number of cavities and linac passes, $N=1$ and $M=2$. A particle will encounter the cavity twice: first when accelerating (encounter $1$), and secondly when decelerating (encounter $2$). Unlike in the case of $A$ and $B$, here $\phi_{\text{in},2}$ is affected by the original cavity phase choice,
\begin{equation}\label{eq:phiin2a}
\begin{aligned}
    \phi_{\text{in},2} &= \phi_{\text{in},1} + \omega (T_{1} + t_\text{pair}) \\
    &=\phi_{\text{out},1} + \omega t_\text{pair},
\end{aligned}
\end{equation}
where $t_\text{pair}$ is the amount of time spent over the return loop between the first encounter's exit and the second's entrance. If the cavity is symmetrically built and the proper $t_\text{pair}$ is found to satisfy the linear sequence symmetry condition from Eq.~(\ref{eq:inputphase}), the 1-cavity ERL will become symmetric. Substitute this condition into Eq.~(\ref{eq:phiin2a}) and set $\phi_{\text{in},B}=\phi_{\text{in},2}$, 
\begin{equation} \label{eq:tpair-2p1c-ERL}
\begin{aligned}
   \omega t_\text{pair} &=  -2 \phi_{\text{out},1} = -2(\phi_{\text{in},1} + \omega T_1) &[\text{odd}]\\
   \omega t_\text{pair} &= \pi-2 \phi_{\text{out},1} = \pi -2(\phi_{\text{in},1} + \omega T_1). &[\text{even}]
\end{aligned}
\end{equation}
This is the necessary $t_\text{pair}$ for symmetry.

\subsection{ERL: 1 Cavity, $M$ Passes}
We now extend the symmetry to an ERL with one cavity, but $M\geq 2$ passes. Let $m$ represent the pass index, where $1\leq m \leq M$. The $m^{th}$ encounter of the cavity must hold a symmetric phase relation with the $(M-m+1)^{th}$ encounter, in the manner of Eq.~(\ref{eq:inputphase}). \par 

In the form of Eq.~(\ref{eq:tpair-2p1c-ERL}), $t_{\text{pair},m}$ designates the time spent on the return loop from the $m^{th}$ encounter exit to the $(M+1-m)^{th}$ entrance. For each $m$, this time between pairs must be correctly related to the phase of the $m^{th}$ encounter by Eq.~(\ref{eq:tpair-2p1c-ERL}). This is done by adjusting the time spent in the $m^{th}$ return loop, $t_{\text{loop},m}$, between the exit of the $m^{th}$ encounter and the entrance of the $(m+1)^{th}$ encounter. \par

The time between cavity pairs, $t_{\text{pair},m}$, includes the times $T_{j}$ spent within the cavity of encounter $j$, for $j=(m+1)$ through $(M-m)$, as well as the in-between $t_{\text{loop},k}$ times of flight, where $k=m$ through $(M-m)$. Therefore,
\begin{equation}\label{eq:tpairm}
   t_{\text{pair},m} = t_{\text{loop},\frac{M}{2}}+2 \sum_{k=m}^{\frac{M}{2}-1} (t_{\text{loop},k} +  T_{(k+1)}),
\end{equation}
where $T_k$ = $T_{(M-k+1)}$ and $t_{\text{loop},k} = t_{\text{loop}, (M-k)}$ due to symmetry. The intermediate loop times become,
\begin{equation}\label{eq:tloopm}
    t_{\text{loop},m} = 0.5(t_{\text{pair},m} - t_{\text{pair},(m+1)}) - T_{(m+1)},
\end{equation}
for $m<\frac{M}{2}$. The highest energy loop has $t_{\text{loop},\frac{M}{2}}=t_{\text{pair},\frac{M}{2}}$. With the full $t_{\text{pair},m}$ expansion from Eq.~(\ref{eq:tpair-2p1c-ERL}), this leads to the loop times,
\begin{equation}\label{eq:tloopm-inoutphase}
\begin{aligned}
    \omega t_{\text{loop},m} = &\phi_{\text{out},(m+1)} - \phi_{\text{out},m} - \omega T_{(m+1)} \\
    = &\phi_{\text{in},(m+1)} - \phi_{\text{out},m}.
\end{aligned}
\end{equation}
This can also be expressed in relative phases, $\phi_{\text{in},j}^v=\phi_{\text{in},j}-\hat\phi_j^v$ for any encounter $j$. In the form of Eq.~(\ref{eq:phiin-gen}), if the particle is ultra-relativistic, then all phase offsets are identically $\hat\phi_{j}^v=\hat\phi^c$. Then, $t_{\text{loop},m}$ from Eq.~(\ref{eq:tloopm-inoutphase}) can be written explicitly,
 \begin{equation}
 \begin{aligned}
     \omega t_{\text{loop},m} = \phi_{\text{in},(m+1)}^c - \phi_{\text{out},m}^c.
 \end{aligned}
 \end{equation}
The phase difference simply accumulates during the travel time if the reference phases $\hat\phi_{j}^v$ do not change between cavity encounters.

\subsection{ERL: $N$ Cavities, $M$ Passes}
Consider a large ERL with $N$ cavities and $M$ passes. As in the 1-cavity ERL, $\phi_{\text{in},mn}$ of all cavity encounters $mn$ after the first pass are dependent on the initial encounter phase, $\phi_{\text{in},1n}$, where again $1\leq m\leq M$ and $1 \leq n\leq N$. Drift pipes between cavities $mn$ and $m(n+1)$ increase particle times by $t_{\text{drift},mn}$, while return loops have time of flight $t_{\text{loop},m}$. The particle is injected into the ERL at a time $t=0$, and it enters cavity $n$ on pass $m$ at the input time $t_{\text{in},mn}$. \par 

Let $t_{\text{total}}$ represent the full time that a beam would take to travel through the symmetric ERL. For brevity, $n'=N-n+1$ and $m'=M-m+1$ will designate the cavity and pass indices for the cavity encounter that is symmetric to the one with $n$ and $m$, where we let $n',m'$ refer to the earlier encounter and $n,m$ to the later encounter, \textit{i.e.} $m>m'$. As in the example of two successive cavities, the voltage of $n$ and $n'$ must satisfy $V_n = V_{n'}$ to permit mirrored cavity interactions during symmetric acceleration and deceleration. If $N$ is odd, then $n=n'$ for the central cavity, but this case can be treated identically to all other paired cavities. \par
\begin{figure}
\includegraphics[width=\linewidth]{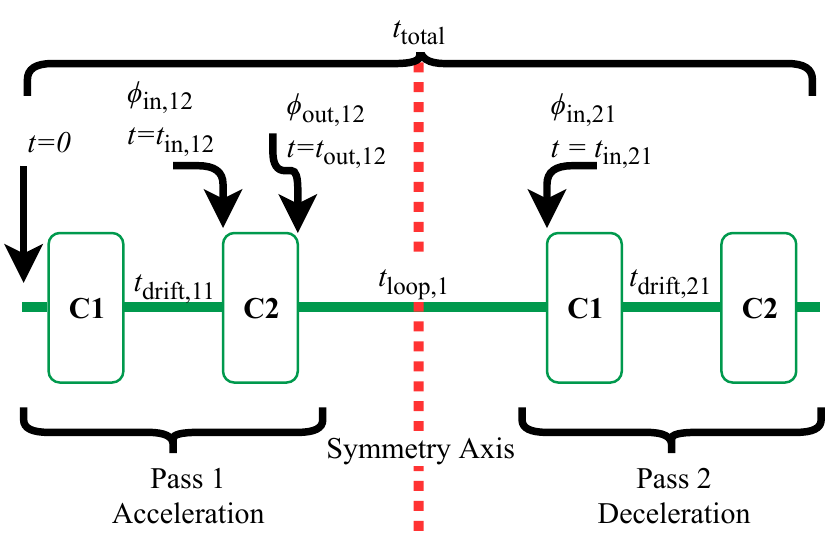}
\caption{Cavity encounter sequence for a 2-pass ERL with $N=2$ physical cavities, denoted C1 and C2, and cavity phase inputs $\phi_{\text{in},mn}$ and outputs $\phi_{\text{out},mn}$ associated with corresponding input/output times $t$. The total time spent in the ERL, $t_{\text{total}}$, is represented by the length of the horizontal line. This schematic can be extended to general multi-pass ERLs by inserting additional cavities or return loops on both sides of the dotted symmetry axis.}
\label{fig:2pass-unwrapped}
\end{figure}
If $\phi_{\text{in},mn}$ is known, use it to define the effective initial cavity phase. Let $\phi_{0,n}$ be the phase seen by a particle that enters cavity $n$ at time $t=0$,
\begin{equation}\label{eq:phi0n}
\begin{aligned}
    \phi_{0,n} &= \phi_{\text{in},mn} - \omega t_{\text{in},mn} \\
    &= \phi_{\text{out},mn} - \omega t_{\text{out},mn},
\end{aligned}
\end{equation}
where the beam exits the cavity at time $t_{\text{out},mn}=t_{\text{in},mn}+T_{mn}$. The phase of the $n^{th}$ cavity must satisfy Eq.~(\ref{eq:inputphase}) to reverse the primed cavity acceleration. Substitute $\phi_{\text{in},mn}$ and $\phi_{\text{out},m'n'}$ from Eq.~(\ref{eq:phi0n}) into the symmetry conditions in Eq.~(\ref{eq:inputphase}) to solve for the unknown $\phi_{0,n}$ in terms of time and known, primed quantities,
\begin{equation}\label{eq:fcav-phisym-deriv}
\begin{aligned}
    \phi_{\text{in},mn} &=  - \phi_{\text{out},m'n'} &[\text{odd}] \\
    \phi_{0,n} + \omega t_{\text{in},mn} &= - (\phi_{0,n'} + \omega t_{\text{out},m'n'})\\
    \phi_{0,n} =&  - \phi_{0,n'} - \omega ( t_{\text{out},m'n'} +t_{\text{in},mn}) \\
     \phi_{0,n} =& \text{ } \pi - \phi_{0,n'} -\omega ( t_{\text{out},m'n'} +t_{\text{in},mn}). &[\text{even}]
\end{aligned}
\end{equation}
When the ERL is symmetric, any point in time from Fig.~\ref{fig:2pass-unwrapped} can be found by stepping forward from $t=0$ or backwards from $t=t_\text{total}$. Cavity entrance times follow this principle,
\begin{equation}\label{eq:fcav-phisym-ttot}
 t_{\text{in},mn} = t_{\text{total}} - t_{\text{out},m'n'}.
\end{equation}
If $t_\text{total}$ is taken as a known parameter, we can relate the primed and unprimed initial phases. Substituting Eq.~(\ref{eq:fcav-phisym-ttot}) into Eq.~(\ref{eq:fcav-phisym-deriv}), the $\phi_{0,n}$ constraints are,
\begin{equation}\label{eq:phisym-phicondition}
\begin{aligned}
 \phi_{0,n} &=   - \phi_{0,n'} - \omega t_{\text{total}} &[\text{odd}]\\
 \phi_{0,n} &=  \pi - \phi_{0,n'} - \omega t_{\text{total}}. &[\text{even}]
\end{aligned}
\end{equation}
To ensure that dependent phases take the proper values in passes $m>1$, the parameter $t_{\text{total}}$ must accurately describe the total time that a beam will spend in the ERL. Since the central loop crosses the symmetry axis of the unwrapped ERL, it is easiest to get a correct $t_{\text{total}}$ by setting the length of loop $\frac{M}{2}$,
\begin{equation}\label{eq:phisym-t4condition}
 t_{\text{loop},\frac{M}{2}} = t_{\text{total}} - 2 t_{\text{out},\frac{M}{2}N},
\end{equation}
where $t_{\text{out},\frac{M}{2}N}$ is the full duration that the beam spends accelerating, from particle injection to exiting the last cavity of the highest-energy pass. \par 
When adjusting the cavities of a physical ERL, it is typical to only have control over the velocity-dependent $\phi_{\text{in},1n}^v$ relative input phases during the first pass. Symmetry conditions for direct ERL control require rearranging Eq.~(\ref{eq:phisym-phicondition}) to solve for the first-pass $\phi_{\text{in},1n}^v$ relative phases,
\begin{equation}\label{eq:phisym-phicondition-v}
\begin{aligned}
    \phi_{\text{in},1n}^v &= -\phi_{\text{in},1n'}^v - \hat\phi_{1n'}^v- \hat\phi_{1n}^v- \omega t_\text{total} &[\text{odd}]\\
    \phi_{\text{in},1n}^v &= \pi-\phi_{\text{in},1n'}^v - \hat\phi_{1n'}^v- \hat\phi_{1n}^v- \omega t_\text{total}. &[\text{even}]
\end{aligned}
\end{equation}
If the particle is ultra-relativistic, then paired cavities $n$ and $n'$ have the same phase offset $\hat\phi_{n'}^c$ in all passes, and Eq.~(\ref{eq:phisym-phicondition-v}) simplifies,
\begin{equation}\label{eq:phisym-phicondition-c}
\begin{aligned}
    \phi_{\text{in},1n}^v &= -\phi_{\text{in},1n'}^c - 2 \hat\phi_{n'}^c- \omega t_\text{total} &[\text{odd}]\\
    \phi_{\text{in},1n}^v &= \pi-\phi_{\text{in},1n'}^c - 2 \hat\phi_{n'}^c- \omega t_\text{total}. &[\text{even}]
\end{aligned}
\end{equation}
For complete ERL symmetry, all $\phi_{0,n}$ with $n\geq n'$ must follow Eq.~(\ref{eq:phisym-phicondition}). The $\frac{M}{2}$ flight time must also follow Eq.~(\ref{eq:phisym-t4condition}). Intermediate return loops may then be set at any reasonable time of flight without disrupting this symmetry. \par 

\section{Cavity Models for ERL Use}
The symmetry conditions and objectives are tested in ERL simulations that use one of the following cavity representations to calculate particle energy and time of flight interactions. Each model is an approximation the real CBETA cavities, which have 7 cells with an elliptical geometry \cite{designreport}.
\begin{itemize}
 \item \textit{Thin Lens Cavities (TL).} Cavities are infinitely thin delta-function energy kicks.
 \item \textit{Ultra-relativistic Cavities (UR).} Particles are treated with ultra-relativistic flight time and energy change formulas inside multi-cell pillbox cavities.
 \item \textit{Finite Time-tracked Cavities (FT).} Cavities are single-cell pillboxes where non-ultrarelativistic particle behavior is considered.
 \item \textit{Runge Kutta Cavities (RK).} Non-ultrarelativistic particle time and energy effects are calculated via integration through multi-cell pillbox cavity fields.
\end{itemize} 
All models track particle time and energy within cavities with ideal pillbox cell shapes. The first three variants are tested in custom Mathematica programs \cite{mathematica}. The latter, which is used as the accuracy reference for the other three, is modeled in the Bmad accelerator toolkit \cite{bmad}. Each model is compatible with the symmetric and reduced symmetric ERL objective systems from Table~\ref{table:freedom-objectives}. \par 

\subsection{Thin Lens (TL) Cavity Model} 
As a first-order approximation, RF cavities can be modeled as a delta-function acceleration over infinitesimal time and distance. Each cavity imparts an instantaneous energy kick consistent with an on-axis beam in an infinitely thin, pillbox-shaped cavity,
\begin{equation}\label{eq:ethin}
 \Delta E_{\text{TL}} = qV \cos(\phi_{\text{in}}^c),
\end{equation}
where the relative input phase, $\phi_{\text{in}}^c$, is used in the same form as Eq.~(\ref{eq:deltaE-A}) due to the negligible cavity transit time. The voltage $V$ describes the maximum possible energy gain of a $q$-charged particle, at the relative input phase $\phi_{\text{in}}^c = 0$. Thin lenses have the position and entrance time of the midpoint of a typical finite-length cavity. Since these cavities have zero length, and effectively zero cells, symmetry in time must be determined using the phase constraint from Eq.~(\ref{eq:inputphase}) for even-cell cavities. \par 
Drift pipes affect the particle time according to particle velocity ($t_{\text{drift}} = \frac{L_{\text{drift}}}{v(E)}$). The pipes on either side of a thin lens are extended by a distance that compensates for the missing length of the thin cavity, thereby keeping the same overall linac dimensions as a more realistic system. \par 

\subsection{Ultra-relativistic (UR) Finite Cavity Model}
The thin lens model neglects the length of a physical cavity. For a better approximation of ERL interaction, we modify the TL equation to model beam energy and time behavior as ultra-relativistic within a cavity of nonzero length. Outside of a cavity, the beam still has typical energy-dependent velocity. For example, if a particle of charge $q$ travels through a 7-cell $\pi$-mode cavity ($n_{\text{cell}}=7$) of length $L=(n_\text{cell} \pi c)/ \omega$ and voltage $V$, then time and energy are modeled as,
\begin{equation}\label{eq:eultra} 
\begin{aligned}
 \Delta E_{\text{UR}}&= qV \cos(\phi_{\text{in}}^c)\\
 T_{\text{UR}} &= \frac{L}{c}=\frac{n_{\text{cell}} \pi}{\omega}.
\end{aligned}
\end{equation}
This energy change and time of flight across the cavity are only accurate for a particle with speed $c$. The general energy gain of any traveling particle with original energy $E_\text{in}$ would follow a more complex relation, and this relationship is further explored the FT model. \par 

\subsection{Finite Time-Tracked (FT) Cavity Model}
The UR model assumes that the electron beam travels at the speed of light, which is not true for a typical MeV-order ERL injection energy. We take this into account by approximating that the particle spends the first half of the cavity length with the initial velocity and half the distance with its final velocity:
\begin{equation}\label{eq:FTtime}
T_\text{FT}= \frac{L}{2}\bigg(\frac{1}{v_\text{in}}+\frac{1}{v_\text{out}}\bigg) ,
\end{equation}
with velocity calculated as,
\begin{equation}
\frac{1}{v} = \frac{1}{c} \sqrt{1+\bigg(\frac{mc}{p}\bigg)^2},
\end{equation}
we then find the momentum,
\begin{equation}\label{eq:p-out}
p_{out} = p_{in} + \frac{q}{\omega} E_\text{in} [ \cos(\omega T_\text{FT} + \phi_\text{in}) - \cos(\phi_\text{in}) ].
\end{equation}
Inserting $T_\text{FT}$ into Eq.~(\ref{eq:p-out}) leads to a transcendental equation for $p_\text{out}$ that can be solved numerically, from which the final energy is obtained by $E_\text{out} = c\sqrt{(mc)^2 + p_\text{out}^2}$. The energy difference is then,
\begin{equation}\label{eq:FTenergy}
   \Delta E_\text{FT} = c\sqrt{(mc)^2 + p_\text{out}^2} - E_\text{in},
\end{equation}
where $E_\text{in}$ is the original particle energy when it enters the cavity.

\subsection{Runge Kutta (RK) Cavities}
In the Bmad accelerator toolkit \cite{bmad}, cavities can be tracked with the \textit{Runge Kutta 4} algorithm. Particle energy and position are found via integration through the electric and magnetic fields of the cavity. If the particle travels through the exact center of the cavity (transversely on-axis), then it will only experience a longitudinal electric field, $\mathscr{E}_s$. If the cavity is an ideal pillbox, and the Bmad \textit{auto-phase} calculation is deactivated (direct $\phi_{\text{in}}$ control), then Bmad models the longitudinal $s$ component of standing wave fields as,
\begin{equation}\label{eq:efield-bmad}
\begin{aligned} 
    \mathscr{E}_s &= \frac{2 V}{L} \sin(ks)\sin(\omega (t-t_\text{in}) + \phi_{\text{in}}), \\
\end{aligned}
\end{equation}
where the particle encounters voltage $V$ over the pillbox length $L=\frac{\pi c}{\omega}$ for RF frequency $\omega$, the speed of light is $c$, and the wave vector is $k=\frac{c}{\omega}$ \cite{bmad}.\par 
Note that for a pillbox run in the fundamental harmonic, $k=0$. The field in Eq.~(\ref{eq:efield-bmad}) is effectively the first harmonic of a pillbox cavity. Since first harmonic fields are anti-symmetric about the center of a cell, but CBETA cavities operated in the fundamental harmonic should be symmetric about the cell center, the typical first harmonic spatial and time dependencies are shifted by $\frac{\pi}{2}$ to create a center-symmetric field pattern. This modified first harmonic field models the physical field in elliptical cavities slightly better than the fundamental pillbox mode would.\par 
Cavities with multiple cells are represented as series of consecutive pillboxes stacked end-to-end. The particle enters the first cell with input phase $\phi_\text{in}$; the output phase of the first cell becomes the input phase for the second cell, the second output feeds into the third input, and so forth. This setup approximates the effect of a $\pi$-mode cavity, although it does not perfectly account for non-ultrarelativistic time dependence between different cells.\par 
When modeling symmetry in CBETA with RK cavities, an on-axis particle is tracked through a 7-cell RK pillbox using a 1D \textit{Runge Kutta} algorithm. 

\subsection{Cavity Model Accuracy}
\begin{figure*}
\includegraphics[scale=0.4]{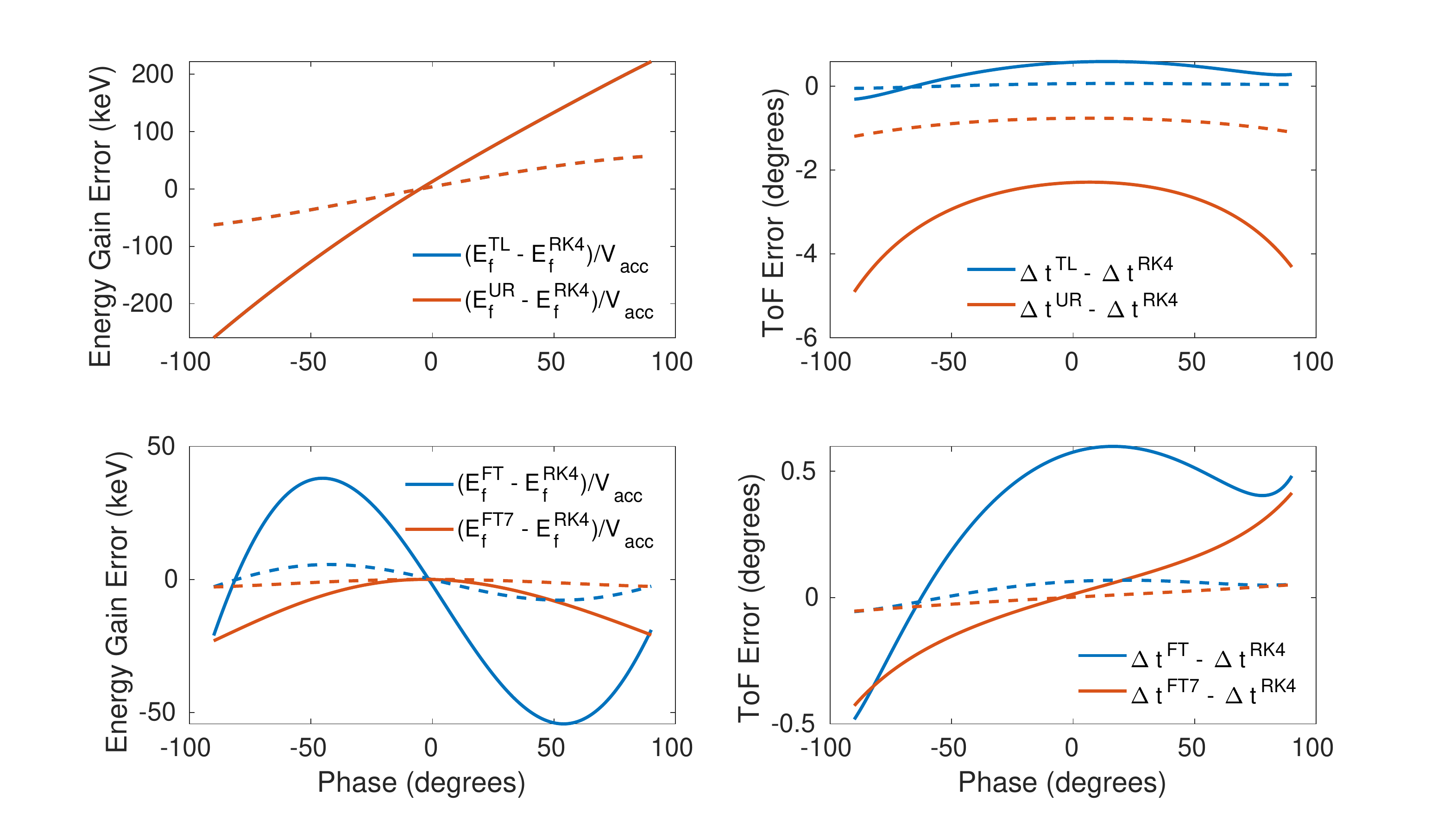}
\caption{Performance of TL, UR, FT, and FT7 cavity models with respect to a RK reference cavity at an incident beam energy of 6~MeV (solid lines) or 12~MeV (dashed lines) and maximum energy gain of 6~MeV. For cavities with length shorter than 7 cells (TL and FT), drift pipes have been added to either side to compensate for the length difference between the models.}
\label{fig:modelcomparison}
\end{figure*}
Physical CBETA cavities have 7 cells with an elliptical geometry. For the purposes of this study, we will approximate these as 7-cell pillbox stacks. To determine how well the mathematical TL, UR, and FT approximations model a real 7-cell pillbox system, we compare them to the most rigorous model: a RK cavity. The RK model is the most realistic in this regard because it directly integrates the particle through the electric fields, while the other models use approximations to determine the travel time and particle energy profiles. \par 
For this comparison, we introduce the FT7 model, which represents a particle moving through 7 identical FT pillbox cavity cells. This is an extension of the FT model, which in its original form describes a single pillbox. The FT7 behavior is included as a demonstration of the difference in accuracy between 1-cell or 7-cell time and energy effects, but FT7 is not used in any subsequent full ERL models.\par

In Fig.~\ref{fig:modelcomparison}, an on-axis particle with an energy of 6~MeV or 12~MeV at the cavity entrance is tracked through TL, UR, FT, FT7, and RK models. Cavities are designed with the CBETA frequency $\omega=(2\pi \cdot 1.3$ GHz$)$, and one cell has length $\frac{\pi c}{\omega}$. Cavities with an active region shorter than 7 cells, such as the zero-length TL model or the 1-cell FT model, are extended to a comparable 7-cell length by adding symmetric drift pipe extensions to either side of the active cavity. Time of flight is then measured from $s=0$ to $s=L$, where $L=\frac{7\pi c}{\omega}$ is the length of a 7-cell cavity. In all models, phase is defined as the $\phi_\text{in}$ at the beginning of the cavity itself, not the drift pipe extensions. \par 
The TL and UR models have an identical, substantial departure from RK energy gain, with errors on the order of 100~keV. The UR time of flight is also the least accurate, due to its assumption of particle $v=c$ within the entire cavity; the TL and FT have near-identical time of flight behavior because the extended drift pipes on either side of the TL cavity create a similar time-average effect as the FT time-average calculation in Eq.~(\ref{eq:FTtime}). In all models, the more relativistic 12~MeV particle experiences less departure from the RK reference than the 6~MeV; the cavity performances converge for particles of higher energy. \par 
Of the models considered, FT7 is most accurate to the RK results, and therefore the closest to expected physical cavity performance. If accuracy to physical systems is the primary concern, then the FT7 should be used; however, if simulation speed is also important, we should consider a trade-off between the desired criteria. The FT7 model requires the simulator to numerically solve 7 sets of Eq.~(\ref{eq:FTtime}) and Eq.~(\ref{eq:FTenergy}), which results in a more time-consumptive calculation than the original FT model. This is not ideal, particularly when a numerical optimization may require thousands of iterations to converge on a solution with the desired objective precision. \par 
For our purposes, the differences in time or energy accuracy between the FT and FT7 models are negligible (Fig.~\ref{fig:modelcomparison}) compared to the increase in simulation speed. In the interests of running a large number of simulations within a reasonable program run-time, the full ERL models in this study consider only the TL, UR, FT, and RK cavities.

\section{CBETA Model Solutions}
The TL, UR, and FT models are simulated with Mathematica scripts, while RK uses the Bmad simulator. Each model ERL is constructed with CBETA-specific parameters: $M=8$ passes, $N=6$ cavities, and the entrances of neighboring cavities are placed $1.41$ m apart. All non-cavity components in the ERL are represented by drift pipes of corresponding lengths. An ideal, on-axis 6~MeV electron beam with zero transverse offset, longitudinal bunch length, or energy spread is injected and reaches a target energy of 150~MeV after 4 accelerating passes. After traveling through the highest energy return loop, the beam decelerates to 6~MeV over the remaining passes. The model cavities use a 1.3 GHz first-harmonic frequency that corresponds with the CBETA design. \par 
The ERL model is made symmetric using the phase conditions from Eq.~(\ref{eq:phisym-phicondition}) and Eq.~(\ref{eq:phisym-t4condition}). Inactive input parameters, \textit{i.e.} the intermediate loop times of flight and all cavity voltages, are manually set at pre-determined values to provide near-maximum acceleration. Intermediate loop times are slightly more than 343 RF periods long, with variations depending on the active length of the cavity model. Voltages are set to achieve an average energy gain of approximately 6~MeV per cavity. For CBETA, pre-optimization settings are chosen such that beam energies fall within $\pm 1$~MeV of the four return loop design energies (42, 78, 114, and 150~MeV).\par 
After pre-optimization settings are found, the degrees of freedom are varied to optimally satisfy the objective functions from Eq.~(\ref{eq:eload}) and Eq.~(\ref{eq:eloop}). Initial phases of cavities 1-3 and the anticipated total time are varied until the objectives, loop 4 energy and cavity 1-3 load, fall within machine precision of their design targets ($E_{\text{loop},4}\rightarrow 150$~MeV, $E_{\text{load},n}\rightarrow 0$). This forms a 4-by-4 reduced optimization system, as shown in Table~\ref{table:freedom-objectives}. The system is numerically solved by both Mathematica and Bmad. The former uses Newton's method, while the latter uses Levenberg-Marquardt differential optimization \cite{lm}.  \par 

\begin{table}
\begin{center}
\begin{tabular}{ |c|c|c|c|c| }
 \hline
 Objective & TL & UR & FT & RK\\
 (\textmu eV) & & & & \\
 \hline

 $\Delta E_{\text{loop},4}$ &  37.1039 & -43.8690 & -72.3600 & 851488\\
 $E_{\text{load},1}$ & -28.7071 &   5.3048 &  20.7573 & -267.860\\ 
 $E_{\text{load},2}$ &  -3.9563 &   7.3761 &  20.6791 & -31.1397\\ 
 $E_{\text{load},3}$ &  -1.2033 &  28.7816 &  30.5586 & -84.9664\\ 
 $E_{\text{load},4}$ &   0.7078 & -28.5134 & -30.8864 & 64570.0\\ 
 $E_{\text{load},5}$ &   2.5891 &  -7.2699 & -21.6011 & 67490.5\\ 
 $E_{\text{load},6}$ &  26.5380 &  -5.6755 & -21.3590 & 76797.8\\ 
 \hline
 \hline
 Energy  & - & - & - & - \\
 (MeV) &  &  &  &  \\
 \hline
 $E_{\text{loop},1}$ &  41.9300 &  42.0047 &  42.0157 & 42.1801\\ 
 $E_{\text{loop},2}$ &  77.9808 &  78.0055 &  78.0167 & 78.2023\\ 
 $E_{\text{loop},3}$ &  114.039 &  114.004 &  114.011 & 114.226\\ 
 $E_{\text{loop},4}$ &  150.000 &  150.000 &  150.000 & 150.000\\
 $E_{\text{loop},5}$ &  114.039 &  114.004 &  114.011 & 114.226\\ 
 $E_{\text{loop},6}$ &  77.9808 &  78.0055 &  78.0167 & 78.2023\\ 
 $E_{\text{loop},7}$ &  41.9299 &  42.0047 &  42.0157 & 42.1801\\ 
 $E_{\text{loop},8}$ &   6.0000 &   6.0000 &   6.0000 &  6.0000\\
 \hline
 \hline
 Input & - & - & - & - \\
 \hline
 $\phi_{0,1}$ ($^\circ$) & -16.5389 &   0.0087 &  1.2841 & 0.3851\\
 $\phi_{0,2}$ ($^\circ$) & -47.6317 & -40.4940 & -39.6981 & -41.5841\\
 $\phi_{0,3}$ ($^\circ$) & -87.7461 & -72.3757 & -78.0626 & -83.5456\\
 $t_{\text{total}}$ (\textmu s) & 2.15392 & 2.15430 & 2.15431 & 2.15429\\
 \hline
 $t_{\text{loop},1}$ (\textmu s)& 0.26418 & 0.26456 & 0.26456 & 0.26456\\
 $t_{\text{loop},2}$ (\textmu s)& 0.26518 & 0.26456 & 0.26456 & 0.26455\\
 $t_{\text{loop},3}$ (\textmu s)& 0.26418 & 0.26456 & 0.26456 & 0.26456\\
 $qV_n$ (MeV) & 6.0500 & 6.0500 & 6.0500 & 6.0500 \\
 \hline

\end{tabular}
\end{center}
\caption{Peak loop energy and cavity load objectives after optimization, resulting beam energy during each loop, and the associated best input settings (phase, total time, loop time of flights, and cavity voltage) after numerical optimization of $\phi_0$ and $t_\text{total}$.}
\label{table:solutions}
\end{table}

During optimization, it is important to check that the optimizer does not select an unphysical time. For example, $t_\text{total} < \sum( t_{\text{loop},m})$ would indicate a negative amount of time spent in the $\frac{M}{2}^{th}$ return loop. If the cavities have a $2\pi$-periodic phase dependence, an unphysical optimized time can be corrected by adding integer multiples of $\frac{2\pi}{\omega}$ until a positive central loop time of flight is found. \par
Optimized values of objective functions, defined in Eq.~(\ref{eq:eload}) and Eq.~(\ref{eq:eloop}) as the differences between modeled load or energy and respective targets, are provided for each model in Table~\ref{table:solutions}. The degrees of freedom and fixed input values are also present.\par 
The optimal $\phi_0$ and $t_\text{total}$ settings for UR, FT, and RK models correspond well, but the optimized TL phase solutions deviate by around 10-15$^\circ$ by model. This may be because $\phi_\text{in}$ for TL cavities is measured at the start of the thin lens itself, which is located in the center of a nonzero-length UR, FT, or RK cavity. Interestingly, the UR, FT, and RK solutions agree to within 2$^\circ$ for the first two cavity phases, but they differ more significantly in the third phase. \par 
The slight difference in intermediate $t_\text{loop}$ return time of flights may also be related to this phenomenon, as the values were chosen for each model individually to achieve pre-optimized beam energies close to the design energies. \par
The three Mathematica models arrive at approximately machine precision on cavity load and central loop energy objectives, as defined in Eq.~(\ref{eq:eload}) and Eq.~(\ref{eq:eloop}). For comparison, the typical beam energy falls on the order of 10-100~MeV, and the objective offsets are on the order of 10-100 \textmu eV. \par 
The 2-3 order of magnitude lower precision on the RK solution is a consequence of the Bmad numerical optimization algorithm used; the convergence of the \textit{lmdif} Bmad optimizer on a solution depend highly on the initial parameter values, objective weights, and step size chosen for optimization. The larger values of the load 4-6 objectives in the RK model, which were not explicitly used in optimization, appear because the RK model setup only achieves about 0.1~eV precision between symmetric accelerating and decelerating energies, as opposed to the \textmu eV-scale energy symmetry established in the three Mathematica models. Nevertheless, in this solution, the RK objective function most poorly satisfied (peak energy, $\Delta E_{\text{loop},4}$) is less than 1~eV away from the 150~MeV target. \par 
With the introduction of ERL symmetry in the four models of increasingly complex cavities, the necessary optimization system is reduced to a 4-by-4 set of objective functions and degrees of freedom. Although only half of the cavity loads were used in optimization, the solutions from Table~\ref{table:solutions} indicate that symmetry does result in correspondence between the optimized loads 1-3 and the non-optimized values of cavities 4-6. These solutions therefore show the practical use of ERL symmetry to satisfy objective functions.\par

\subsection{Solution Sensitivity}
Once optimal solutions are identified, we introduce errors to the input parameters to determine the accuracy required for successful ERL operation. The inputs to be varied include all $N$ phases, $N$ voltages, and $\frac{M}{2}$ loop lengths in the system; output functions are the objectives of maximum beam energy and all $N$ cavity loads. For small input permutations about the Table~\ref{table:solutions} solutions, the objective function response is approximately linear; we can therefore speak of an approximately constant slope (sensitivity) of each objective to a single degree of freedom. \par
Objectives must be kept within a certain tolerable range around the ideal values (zero load, 150~MeV peak energy) for the ERL to operate in an acceptable manner. If the tolerable range is small enough with respect to the curvature of the objective function's dependence on the inputs, then the linear response model holds. Assume that all inputs except one, denoted $j$, are set to the optimized values from Table~\ref{table:solutions}. We denote each objective function as $f(j)$ for a particular objective $f$, such that $f(j)=E_{\text{load},n}(j)$ or $f(j)=\Delta E_{\text{loop},\frac{M}{2}}(j)$. The input $j$ is therefore one of $2N+\frac{M}{2}$ total input parameters, and the function $f(j)$ is one of the $N+1$ objectives in consideration.\par 
The objective tolerance, $f_0$, can be divided by that same objective's sensitivity to error in $j$, $\frac{\text{d}}{\text{d} j}f(j)$, to find the maximum error $\Delta j_f$ that the input $j$ can have before $f(j)$ exceeds tolerance. The smallest value in the set of $\{\Delta j_f\}$, for all functions $f(j)$, therefore satisfies all $f_0$ tolerances. 
\begin{equation}\label{eq:sensitivity-individual}
\begin{aligned}
    \Delta j_f &= f_0 \bigg(\frac{\text{d}}{\text{d} j} f(j)\bigg)^{-1} \\
    \Delta j &= \text{min}(\{ \Delta j_f \}).
\end{aligned}
\end{equation}
Input $j$ can be safely varied from its optimized solution until the magnitude of objective function $f(j)$ reaches its tolerance limit. \par 
The CBETA $f_0$ tolerances are chosen as a maximum load of 50~keV per cavity (2 kW for a 40 mA injection current), and 150~keV offset from peak beam energy. If the tolerances were large enough to give nonlinear objective function responses, then a more detailed analysis may be required; however, our 50~keV and 150~keV tolerances are sufficiently small that a linear response model of objective dependencies is valid. We then use Eq.~(\ref{eq:sensitivity-individual}) to determine the $\Delta j$ of each input when only that input is varied from the ideal solution. \par 

\begin{table}
\begin{center}
\begin{tabular}{ |c|c|c|c|c|c|c| } 
 \hline
 Phase ($^\circ$) & TL  & UR & FT & FT$_{12}$ & RK \\
 \hline
 $\phi_{0,1}$ & 1.7339 & 10.0944 & 3.7625 & 4.6298 & 0.4957 \\ 
 $\phi_{0,2}$ & 8.8545 &  7.0988 & 3.3278 & 3.4564 & 0.4776 \\ 
 $\phi_{0,3}$ & 7.3789 &  1.6782 & 2.2341 & 2.0091 & 0.4725 \\ 
 $\phi_{0,4}$ & 7.3845 &  1.6852 & 2.2395 & 2.0170 & 0.4725 \\ 
 $\phi_{0,5}$ & 8.9274 &  7.0731 & 3.2569 & 3.4344 & 0.4776 \\ 
 $\phi_{0,6}$ & 1.7004 &  9.1458 & 3.3749 & 4.3917 & 0.4957 \\ 
 \hline
 \hline
 Voltage (keV) & - & - & -  & - & -\\
 \hline
 $qV_1$ & 38.103 & 37.541 & 37.758 & 37.642 & 37.391 \\ 
 $qV_2$ & 37.433 & 37.557 & 37.757 & 37.719 & 37.609 \\ 
 $qV_3$ & 37.470 & 38.373 & 38.003 & 38.111 & 37.617 \\ 
 $qV_4$ & 37.495 & 38.368 & 37.992 & 38.105 & 37.807 \\ 
 $qV_5$ & 37.524 & 37.550 & 37.732 & 37.708 & 37.851 \\ 
 $qV_6$ & 38.352 & 37.531 & 37.716 & 37.629 & 37.871 \\ 
 \hline
  \hline
 Return (mm) & - & - & -  & - & -\\
 \hline
 Loop 1 & 0.2557 & 0.2389 & 0.3162 & 0.2854 & 0.3277 \\ 
 Loop 2 & 0.3853 & 0.3582 & 0.4704 & 0.4280 & 0.3793 \\ 
 Loop 3 & 0.7730 & 0.7164 & 0.9478 & 0.8558 & 0.4503 \\ 
 Loop 4 & 0.3648 & 0.3582 & 0.4740 & 0.4276 & 0.6045 \\ 
 
 \hline

\end{tabular}
\end{center}
\caption{Error ranges where all objectives, $f(j)$, are within tolerance when only the indicated $j$ parameter is imperfect. Ranges represent $\pm \Delta j$ offsets from ideal settings, as calculated from Eq.~(\ref{eq:sensitivity-individual}). Injected particle energy is 6~MeV in all cases except the FT$_{12}$ column, which represents a FT model with an injector energy of 12~MeV instead of 6~MeV. }
\label{table:sensitivity-indepvars}
\end{table}
In Table~\ref{table:sensitivity-indepvars}, sensitivity is inversely proportional to the tolerable error range of a given parameter. Voltage sensitivity appears similar across all models. The RK model phases are consistently the most sensitive to error, while the TL and UR phase sensitivities vary from high ($\pm1.7^\circ$) to low ($\pm 10^\circ$) based on cavity number. In the TL model, the $\phi_{0,1}$ input may have the highest sensitivity because TL cavities are separated by the longest inter-cavity drifts of all models (1.41~m), and the first acceleration in each pass has the largest impact on these drift times. In contrast, the UR model has shorter drifts (0.60~m) with a fixed time interval of $\frac{7\pi}{\omega}$ within each cavity: the drift time effect from the TL model is less prominent. The high sensitivity in the $\phi_{0,3}$ pair of the UR model results from breaking the phase symmetry condition, Eq.~(\ref{eq:phisym-phicondition}). The neighboring cavities (3 and 4) can influence each others' loads more directly than non-adjacent pairs (1 and 6; 2 and 5), which distribute the broken symmetry effect among themselves and all in-between cavities. The FT and RK model phases fall between these two extremes due to the presence of both energy-dependent transit times and symmetry-breaking phases. \par 
The loop length sensitivity of all models decreases as the particle moves from loops 1 to 3. As a particle moves closer to the optimized center of symmetry in loop 4, the energy impact of symmetrically extending acceleration and deceleration paths (loops 1 and 7; 2 and 6; or 3 and 5) decreases proportional to the number of remaining accelerations. A change in the length of loop 1 affects the timing of acceleration in pass 2, 3, and 4; a change in loop 3 only affects acceleration in the 4$^{th}$ pass. Symmetry is preserved when errors are introduced in loops 1, 2, or 3. However, any error introduced in loop 4 breaks the symmetry, as the central loop length condition from Eq.~(\ref{eq:phisym-t4condition}) is no longer valid. This phenomenon may be responsible for the higher loop 4 error sensitivity in TL, UR, and FT models compared to loops 1-3. \par 
In the FT$_{12}$ column of Table~\ref{table:sensitivity-indepvars}, the injector energy of an FT model ERL is raised from 6~MeV to 12~MeV, the maximum energy target is raised from 150~MeV to 156~MeV, and a solution has been optimized to machine precision on energy recovery and the 156~MeV target. The tolerable error ranges of four phases and two voltages increase, but others decrease. The tolerable error on loop length decreases for this higher injector energy. These results indicate a complex relation between solution sensitivity and injector energy, although further investigation in this subject would be needed to determine any trends. \par 
The individual sensitivity results are useful to establish upper bounds for individual input error. However, in reality, the combined effect from multiple error sources must be considered. Further work is required to study the combined sensitivity. \par
Suppose that we only have control over the input designated $j$. Furthermore, suppose the expected error on every other input $i$, for $i\neq j$, has a known standard deviation, $\sigma_{i0}$. Using statistical error propagation formulas, we can calculate an allowed error range $\sigma_{j_f}$ for the special input $j$, such that the objective $f(|j|\leq \sigma_{j_f})$ is still within tolerance,
\begin{equation}\label{eq:sensitivity-merger}
\begin{aligned}
 \sigma_{j_f} &= \frac{\sigma_{j0}\sigma_{f0}}{\sqrt{\sum_i (\frac{df}{di} \sigma_{i0})^2}} \\
 \sigma_{j} &= \text{min}(\{\sigma_{j_f}\}).
 \end{aligned}
\end{equation}
Error ranges for all objectives $f(j)$ can be combined into the set $\sigma_{j_f}$, which has as many elements as the number of system objectives. The smallest $\sigma_{j_f}$ is the range of $j$ required to satisfy the most restrictive objective; therefore the range required to satisfy all objectives is $\sigma_j=\text{min}(\{\sigma_{j_f}\})$.

Using the same tolerance bounds as for the single-variable errors, we set $\sigma_{f0}=50$~keV for load, and $\sigma_{f0}=150$~keV for the maximum beam energy. In CBETA, the expected $\sigma_{j0}$ fluctuations for phase, voltage, and return loop length inputs are $0.1^\circ$, 600~eV, and $\frac{1}{3}$~mm, respectively. These fluctuation ranges are consistent with the precision of CBETA's existing low-level RF (LLRF) system, as well as that of other state of the art superconducting RF (SRF) linacs \cite{llrf-control-res}. A calculation of combined tolerances using Eq.~(\ref{eq:sensitivity-merger}) gives the Table~\ref{table:sensitivity} error ranges for individual input $j$ parameters with assumed $\sigma_{i0}$ fluctuations on all other settings.

\begin{table}
\begin{center}
\begin{tabular}{ |c|c|c|c|c|c| } 
 \hline
 Input & Expected & TL  & UR  & FT   & RK \\
 Group &  $\sigma_{j0}$ & &  & & \\
 \hline
 $\phi_{0,n}$ ($^\circ$) & 0.1000 & 0.0536 & 0.0507 & 0.0671 & 0.0608 \\ 
 $qV_n$ (keV) & 0.6000 & 0.3222& 0.3040 & 0.4024 & 0.3651 \\ 
 Loop (mm) & 0.3333 & 0.1790 & 0.1689 & 0.2235 & 0.2028 \\ 
 \hline

\end{tabular}
\end{center}
\caption{Error tolerance ranges for cavity phase, voltage, and return loop length, as calculated using Eq.~(\ref{eq:sensitivity-merger}) with expected $\sigma_{j0}$ fluctuations on all input parameters. For a system that can feasibly achieve and maintain operation within pre-defined objective bounds, the input tolerance should be larger than expected $\sigma_{j0}$.}
\label{table:sensitivity}
\end{table}

The acceptable uncertainties listed in Table~\ref{table:sensitivity} represent the stability required from the cavity field regulation system and the resolution required from the path length adjustment system. The CBETA regulation system must keep dynamic effects, such as noise or transient RF field responses, within the $\sigma_{j0}$ limits in order to satisfy the objective tolerances. The combined sensitivity results in Table~\ref{table:sensitivity} indicate that each input $j$ must be regulated about twice as strictly as the existing control precision in order to prevent objective values from exceeding the $\sigma_{f0}$ tolerance bounds. If the modeled sensitivities are representative of real CBETA error sensitivity, these results may indicate a need for improved control systems in order to maintain operation that satisfies our load and energy tolerances.\par

\section{Longitudinal Beam Dynamics}
The modeled solutions and sensitivities thus far have only considered a single-particle, on-axis beam with zero time or energy spread. A more realistic particle distribution is expected to have non-zero dimensions in phase space ($x,p_x,y,p_y,z,p_z$), where all coordinates are relative to the position of the ideal single-particle scenario. By convention, the six-by-six [$R_{ij}$] matrix is defined to map all phase space coordinates from an injected particle into its final position. Since our models only track the particle in the longitudinal phase space, all elements involving ($x,p_x,y,p_y$) coordinates follow the form of an identity matrix. Consider a transversely on-axis beam ($<x,y,p_x,p_y>=\bm{0}$) with finite spread in injection time and energy (momentum). All longitudinal phase space coordinates are defined with respect to an ideal particle with coordinates $z=0$, $\delta=\frac{p_z-p_0}{p_0}=0$. \par 
In the following analysis, we continue to model CBETA as a sequence of cavities and drifts. All drifts add a velocity-dependent time to the particle coordinates. In CBETA, the splitter and recombiner regions are adjusted to minimize the time-energy dependence ($R_{56}$) incurred throughout each FFA return loop. Ideally, CBETA should have $R_{56}=0$ in each loop. In the TL, UR, FT, and RK models, all return loops and inter-cavity drifts are modeled as straight drift pipes, where $t_\text{loop}=\frac{length}{velocity}$. This yields $R_{56}\leq 0.01$ throughout the entire ERL. \par 
The injected beam follows a Gaussian distribution in time and energy, with $\sigma_\text{time}=4$ ps ($\sigma_{z}\approx1.1$~mm) and $\sigma_\delta=5\cdot10^{-4}$. In the desired scenario, an injected beam should achieve minimum $\delta$ spread during its highest-energy pass. This will allow better control over the individual particle energies after acceleration if a fraction of the beam is siphoned off for experimental purposes. Furthermore, to preserve the acceleration-deceleration symmetry found in the optimized ideal particle solutions, it is preferred if the beam has equivalent energy spread in the longitudinal phase space profiles at the injector and stop. \par
\begin{figure}
\includegraphics[width=\linewidth]{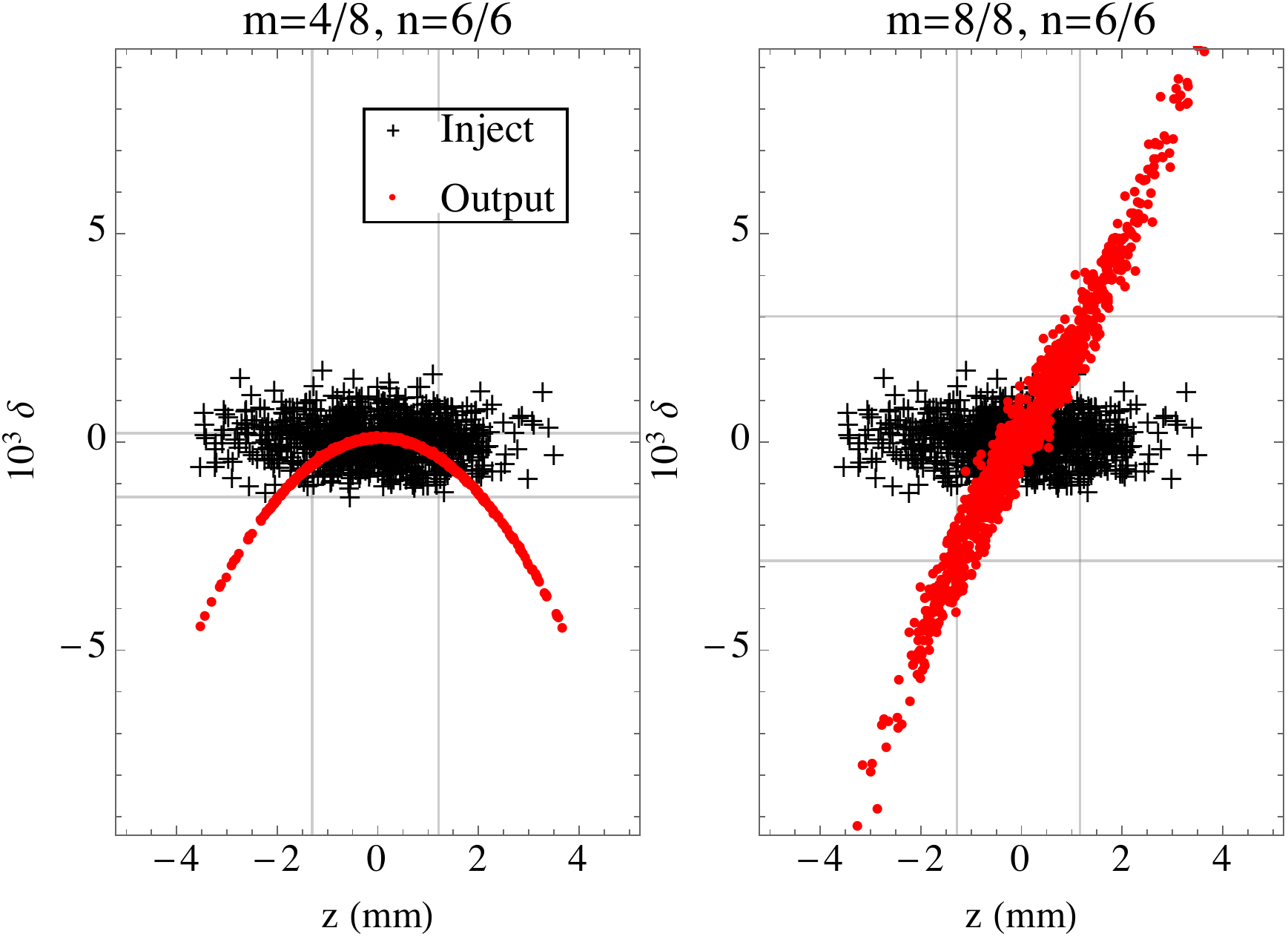}
\caption{Time evolution of 1000-particle Gaussian beam in longitudinal phase space: an initial Gaussian (both plots, overlaid), the beam after complete acceleration (left), and at beam stop (right). The beam is in an FT model ERL with $M=8$, $N=6$, and the ERL is set with ideal particle solutions (Table~\ref{table:solutions}). The ideal particle is defined at coordinates $(z,\delta)=(0,0)$. Vertical and horizontal lines represent one standard deviation of the stated output from the mean $z$ or $\delta$ coordinate.}\label{fig:beam}
\end{figure}

 
When a Gaussian beam is run through an ERL with settings optimized for the ideal particle, the longitudinal phase space profile becomes distinct from the original distribution after a complete, 8-pass ERL circulation (Fig.~\ref{fig:beam}).  \par 
Even in a symmetric ERL, the phase space distributions of a non-ideal beam before injection and after beam stop are not necessarily symmetric. The high energy spread at beam stop may be reduced if more objectives and degrees of freedom are added to the optimization system. A new objective could limit the energy spread of the final beam to a certain tolerable range, but this would also require new degrees of freedom to ensure a useful optimization system. For example, the voltages of paired cavities could be independently varied instead of set to a common value.\par
\begin{figure}
\includegraphics[width=\linewidth]{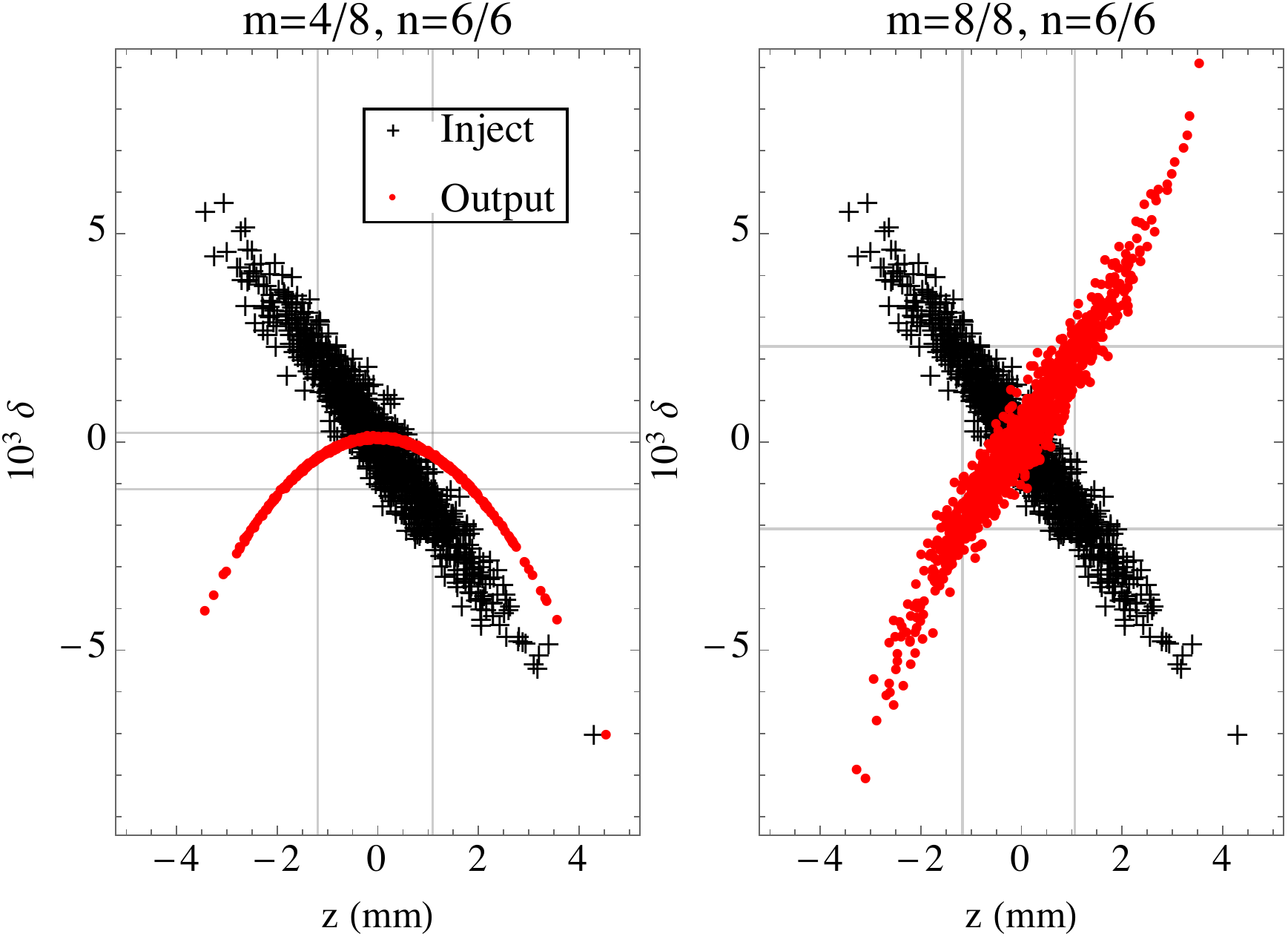}
\caption{Tilted 1000-particle beam in longitudinal phase space for the FT model, shown after full acceleration (left) and deceleration (right). The input distribution is a tilted Gaussian: each particle's original Gaussian $\delta$ coordinate is offset according to the matrix slope from Eq.~(\ref{eq:slopeA}) and Table~\ref{table:beamtilts}; due to the relatively large initial beam size, the edges of the final beam do extend beyond the linear regime.}\label{fig:beamtilt-matrix}
\end{figure}

 
An easier solution to the phase space asymmetry involves adjustment of the injected particle distribution. Since the ERL converts a flat Gaussian $z,\delta$ distribution into a diagonal one, it is reasonable to suspect that, if we inject a beam with initial tilt in the opposite direction, a symmetric mirror image of the phase space distribution may emerge after the ERL (\textit{e.g.} Fig.~\ref{fig:beamtilt-matrix}, right). This beam would then have equal energy spreads in injection and stop. Such an injection pattern may also decrease the energy spread during the $M=4$ return loop, which will allow better control of the beam for experimental purposes.\par
\begin{figure}
\includegraphics[width=\linewidth]{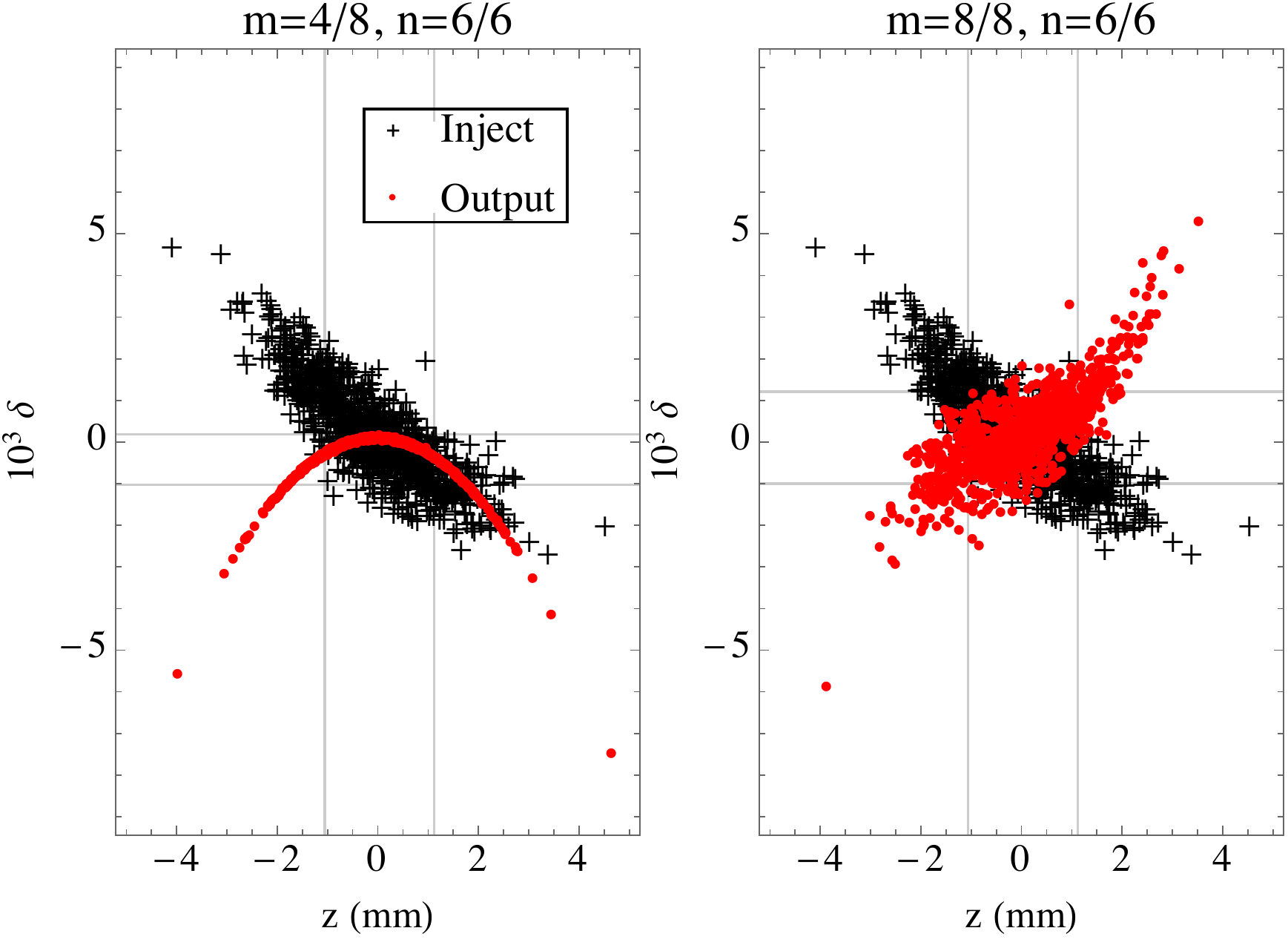}
\caption{Tilted 1000-particle beam in the UR model after full acceleration (left), and at beam stop (right), where the beam tilt is determined by scanning through injector phases until the one that yields the most symmetric input/output $\sigma_\delta$ is found. Note that the center of the final distribution after the ERL is a mirror image of that before entering the accelerator. This symmetry also determines that the slope of the distribution is zero at its center after half the ERL has been traversed. }\label{fig:beamtilt-scan}
\end{figure}

 
The amount of tilt can be adjusted by varying the phase of the final cavity in the injector section. A proper choice of tilted input beam, such as the example in Fig. \ref{fig:beamtilt-scan}, allows equal initial and final $\sigma_\delta$. \par 
To identify the correct input distribution for a symmetric output energy spread, consider the time and energy effects across the complete ERL, from injection to beam stop, as the $5^{th}$ and $6^{th}$ elements of the full coordinate transfer matrix, [$R_{ij}$], \par 
\begin{equation}\label{eq:long-matrix-Rfwd}
\begin{bmatrix}
    z_\text{stop}\\
    \delta_\text{stop}\\
\end{bmatrix}
=
\begin{bmatrix}
    R_{55}       & R_{56} \\
    R_{65}       & R_{66} \\
\end{bmatrix}
\begin{bmatrix}
    z_\text{inj}\\ 
    \delta_\text{inj}\\ 
\end{bmatrix}.
\end{equation}
Elements can be evaluated numerically by injecting test particles at known small ($z_\text{inj},\delta_\text{inj}$) offsets from the ideal particle, which is defined at coordinates ($0,0$), and measuring the final ($z_\text{stop},\delta_\text{stop}$) values at the end of the ERL. We additionally construct an equivalent [$Q_{ij}$] matrix, which describes the longitudinal phase space mapping of a particle that travels backward through the ERL, from injector to beam stop,
\begin{equation}\label{eq:long-matrix-Qbwd}
\begin{bmatrix}
    -z_\text{inj}\\
    \delta_\text{inj}\\
\end{bmatrix}
=
\begin{bmatrix}
    Q_{55}      & Q_{56} \\
    Q_{65}      & Q_{66} \\
\end{bmatrix}
\begin{bmatrix}
    -z_\text{stop}\\ 
    \delta_\text{stop}\\ 
\end{bmatrix}.
\end{equation}
Due to the symmetry between ERL acceleration and deceleration, the mapping of $(z_\text{inj},\delta_{inj)}$ to $(z_\text{stop},\delta_\text{stop})$ must be identical in both Eq.~(\ref{eq:long-matrix-Rfwd}) and Eq.~(\ref{eq:long-matrix-Qbwd}) in order for particles with offsets $(z_\text{inj},\delta_{inj)}\neq (0,0)$ at injection to regain offsets of the same magnitude at the beam stop. Therefore, $Q_{55} = R_{66}, Q_{56} = R_{56}, Q_{65} = R_{65},$ and $Q_{66} = R_{55}$. \par 
If the initial particle has only a small offset from the ideal case, then the ERL mapping can be considered linear, and the two coordinates are linearly related ($\delta=A z$). For the initial and final beam distributions to have equal and opposite tilt, it is required that,
\begin{equation}\label{eq:long-matrix}
\begin{bmatrix}
    z\\
    -A z\\
\end{bmatrix}
=
\begin{bmatrix}
    R_{55}       & R_{56} \\
    R_{65}       & R_{66} \\
\end{bmatrix}
\begin{bmatrix}
    z\\ 
    A  z\\ 
\end{bmatrix}.
\end{equation}
An injected beam with small $\sigma_z$ offset must then have the net linear slope,
\begin{equation}\label{eq:slopeA}
    A=\frac{1-R_{55}}{R_{56}} = -\frac{R_{65}}{1+R_{66}}.
\end{equation}
With this slope $A$, the beam will exit the ERL with the same $\sigma_\delta$ energy spread and an opposite tilt angle as it had at injection. In Table~\ref{table:beamtilts}, a particle with $z=10^{-7}$~m or $\delta=10^{-7}$ offset is used to calculate slopes with Eq.~(\ref{eq:slopeA}). These slopes are used to add a vertical offset to all particles in a Gaussian distribution of ($z,\delta$) particle coordinates (Fig.~{\ref{fig:beamtilt-matrix}}); we then report the tilt amount as the angle of a linear fit of the beam from the positive $z$-axis. \par 

For beams with large $\sigma_z$, the linear matrix transform no longer accurately describes the behavior of particles at the edges of the beam. With this type of beam, the proper tilt pattern is determined by injecting multiple test beams with different injector cavity phases, $\phi_\text{in}$.\par 

A scan for the proper initial beam is conducted by injecting multiple test beams of different longitudinal profiles through the ERL. First, we simulate the pre-ERL injector cryomodule as a single accelerating UR-model cavity with a voltage of $1.5$~MV. A beam with Gaussian energy and time distributions passes through this injector module. This beam is distributed about a pre-injection particle energy and time, $E_0$ and $t_0$, such that after tilting, the central particle will have ideal particle injection properties, $E_\text{inj}=6$~MeV and $t_\text{inj}=0$~s. Pre-injection parameters $E_0$ and $t_0$ are determined by the reverse process of calculating which input energy and time will result in ideal injection parameters after accelerating through the injector module. The tilted beam is sent into the ERL, and its output $\sigma_\delta$ is measured. \par 
We then repeat the process with a new test beam by varying the injector phase in steps of, for instance, $\frac{\pi}{100}$, while using the same pre-injector Gaussian distribution. Once the full range of phases is covered, from 0 to 2$\pi$, we identify the injected beam that yields the most similar $\sigma_\delta$ energy spread in the injected and output distributions, where $\frac{\sigma_{\delta,\text{stop}}}{\sigma_{\delta,\text{inj}}}\rightarrow 1$. The ideal tilt would yield equal energy spread at injection and beam stop, but practical beam results are limited by the resolution of the scan. \par   

\begin{table}
\begin{center}
\begin{tabular}{ |c|c|c|c|c| } 
 \hline
 Matrix & TL & UR  & FT & RK \\
 \hline
  Slope $A$ & 3.2029 & -0.6714 & -1.7748  & -62.3799 \\
Inj. Angle ($^\circ$) & 72.6608 & -33.8772 & -60.6006  & -89.0816\\
 Pass 4 Angle ($^\circ$)& -0.9154 & -0.0042 & -1.0045 & 6.7461  \\
 Stop Angle ($^\circ$) & -71.5540 & 41.2810 & 63.4172  & 89.0879 \\
 $\sigma_{E,\text{inj}}$ (keV) & 21.8924 & 5.3765 & 12.0839  & 418.450 \\ 
 $\sigma_{E,\text{stop}}$ (keV) & 20.5407 & 6.6398 & 13.5756  & 427.277 \\ 
 $\sigma_{\delta,\text{stop}}/ \sigma_{\delta,\text{inj}}$ & 0.9383 & 1.2350 & 1.1234 & 1.0211 \\ 
 \hline
 \hline
 Scan & - & - & -  & - \\
 \hline
  $\phi_{\text{inj}}$ ($^\circ$) & 27.0 & - 7.2 & -18.0 & -90.0 \\
 Inj. Angle ($^\circ$) & 72.1174 & -40.6386 & -64.4857 & -81.5250\\
 Pass 4 Angle ($^\circ$)& 0.3247 & -2.0255 & -4.2896 & 71.3807\\
 Stop Angle ($^\circ$) & -71.7822 & 37.9926 & 60.5429 & 89.4000\\
 $\sigma_{E,\text{inj}}$ (keV) & 20.8529 & 6.7300 & 14.3420  & 44.1964\\ 
 $\sigma_{E,\text{stop}}$ (keV) & 20.5108 & 6.6338 & 12.4481 & 1009.8 \\ 
 $\sigma_{\delta,\text{stop}}/ \sigma_{\delta,\text{inj}}$ & 0.9836 & 0.9857 & 0.8680 & 22.4978 \\ 
 \hline

\end{tabular}
\end{center}
\caption{Beam tilt parameters for a symmetric injector-stop energy spread, found for a 1000-particle CBETA beam ($\sigma_z=1.1$~mm,~$\sigma_\delta=5\cdot 10^{-4}$) with matrices (top), or by scanning through phases of a UR injector (bottom). \textit{Angle} describes the counter-clockwise angle between the positive $z$-axis in longitudinal phase space and a linear fit of the beam. The beam of order $10^{-3}$ in both coordinates is much larger than the $z=10^{-7}$~m,~$\delta=10^{-7}$ offset used to calculate the matrix, resulting in decreased accuracy for matrix tilted values. }
\label{table:beamtilts}
\end{table}
Despite starting with Gaussian distributions of the same mean and standard deviation, the scan and matrix tilting methods yield different optimal linear-fit tilt angles. This is because beams that have passed through an injector cavity experience nonlinear curvature, while artificially tilted Gaussian beams will retain the uniform slope imposed in Eq.~(\ref{eq:slopeA}). For a beam with balanced acceleration and deceleration phase space profiles, injected beams that are tilted to match the Table~(\ref{table:beamtilts}) profiles will most effectively preserve the ERL symmetry from the optimized single-particle settings. \par 
The TL model requires a positive initial tilt angle for the most symmetric input/output energy spread, while UR, FT, and RK models use a negative angle. The opposite tilt needed may be a result of the cavity lengths: while UR, FT, and RK models have 7 cells (odd), TL has 0 effective cells (even).  \par 
The matrix of the RK model has an unexpectedly large $R_{65}$ value of 165~m$^{-1}$, where $\delta_\text{stop}=R_{65}z_\text{inj}$ when only $z_\text{inj}$ is nonzero. In the other models, all matrix elements have magnitudes ranging between 0 and 7 (units of m, m$^{-1}$, or unitless). In the RK model, a particle with a small initial time displacement will incur far larger energy displacement after reaching beam stop. The slope calculation then requires a nearly $90^\circ$ linear tilt angle for tilt symmetry: this is the arrangement with a ratio of time and energy spread, $\frac{\sigma_{z,\text{inj}}}{\sigma_{\delta,\text{inj}}}$, that best matches the ERL matrix. The single UR injector cavity used in the scan is unable to produce the extreme tilt angle needed for the RK model; its most tilted beam output is less than $82^\circ$ from the horizontal, yet this tilt is insufficient to create equivalent energy spreads at injection and beam stop. To generate more tilt, a higher injector cavity voltage would be needed, or multiple injector cavities could be applied. \par

\section{Discussion}
In each of the successive models (TL, UR, FT, RK), the complexity of the cavity time and energy tracking has been increased successively to better model the behavior a realistic cavity. These steps were useful in determining whether the initial time symmetry conditions derived for a thin cavity situation could also be applied to finite-length, non-ultrarelativistic cavities, and finally to cavities that consider a full integrated electric field profile. \par 
For an ideal single particle, symmetry enforcement using specific phase and flight time relations provide optimized solutions well within the tolerable range of cavity power load and peak beam energy objectives. However, the combined sensitivity analysis suggests that our expected instrument fluctuation range ($\sigma_{j0}$) is about twice as large as the error limits required to guarantee fulfillment of all objectives. If optimized solutions from the models are implemented in a system with the existing input control resolution, the actual parameter settings may differ from corresponding optimized values by a larger error margin than needed to satisfy the objectives. As a result, the objectives in such an imperfectly set system may not fall within the desired load and energy target tolerances. The discrepancy between existing and required error ranges indicates a need for better control resolution than the quantities expected for CBETA that we considered in the sensitivity analysis. In practice, this may be achieved by either searching for solutions with lower error sensitivity, or by improvement of the physical systems.  \par 
Applying symmetry to a system such as CBETA allows simplification in the optimization process for ERLs like CBETA. The models in this study considered only cavity, drift pipe, and return loop pipe elements; more complex effects, such as beam optics and transverse dynamics, have not yet been examined. Further work is needed to confirm how representative the pillbox cavity models are of ERL loading and beam dynamics, as well as whether ERL symmetry is feasible for implementation in a system with elements beyond simple RF cavities and drifts. But this lies outside the scope of this paper. Nevertheless, the initial solutions found here indicate that symmetry enforcement is useful for optimizing a reduced set of phase and flight time settings to achieve cavity load and energy objectives. The phases determined here phases will be used during beam commissioning of CBETA.

\section{Acknowledgements}
We thank Dr.~David Sagan for assistance with the Bmad simulation software. We thank Dr.~Christopher Mayes for assistance with the cavity fields. This work was performed through the support of the New York State Energy Research and Development Agency (NYSERDA).

\par
\par


%

\end{document}